\documentclass[12pt,aps,prd,nofootinbib,superscriptaddress]{revtex4-2}

\usepackage{graphicx}
\usepackage{amsmath}
\usepackage{amsfonts}
\usepackage{amssymb}
\usepackage{graphicx}
\usepackage{subfig}
\usepackage{color}
\usepackage{xspace}
\usepackage[colorlinks=true, allcolors=blue]{hyperref}
\usepackage[capitalise]{cleveref}
\usepackage{threeparttable}
\usepackage{multirow}

\newcommand{\GeantFour}{\textsc{Geant4}\xspace}

\begin{document}

%\preprint{}

%\title{Simulation of muonium production from perforated silica aerogel}
\title{Optimization of muonium yield in perforated silica aerogel}


\author{Shihan Zhao}
% \email[]{zhaoshh7@mail2.sysu.edu.cn}
\author{Jian Tang}
\email[]{tangjian5@mail.sysu.edu.cn}
%\thanks{}
\affiliation{School of Physics, Sun Yat-sen University, Guangzhou 510275, China}

\date{\today}

\begin{abstract}
    A muonium consists of a positive muon associated with an orbital electron, and the spontaneous conversion to antimuonium serves as a clear indication of new physics beyond the Standard Model in particle physics.
    One of the most important aspects in muonium-to-antimuonium conversion experiment (MACE) is to increase the muonium yield in vacuum to challenge the latest limit obtained in 1999.
    This study focuses on a simulation of the muonium formation and diffusion in the perforated silica aerogel.
    The independent simulation results can be well validated by experimental data.
    By optimizing the target geometry, we find a maximum muonium emission efficiency of $7.92(2)\%$ and a maximum vacuum yield of $1.134(2)\%$ with a typical surface muon beam, indicating a 2.6 times and a 2.1 times enhancement, respectively.
    Our results will pave the way for muonium experiments.
\end{abstract}

%\keywords{}

\maketitle

\section{Introduction}

Muonium, as a pure leptonic bound state ($\rm M=\mu^+e^-$), is a ideal subject in precision measurement of the Standard Model and a sensitive probe to new physics beyond SM.
Improving the yield of vacuum muonium will make significant contributions in muonium spectroscopy~\cite{Kanda:2020mmc,MuSEUM:2020mzm,Mu-MASS:2021uou,Janka:2022pis,Cortinovis:2023zqi}, precise measurement of muon $g-2$ and electric dipole moment (EDM)~\cite{Iinuma:2011zz,Sato:2021aor,Otani:2022wlj}, muonium antimatter gravity experiment (MAGE)~\cite{Kirch:2014mna,MAGE:2018wxk,Hajdukovic:2019bwd}, and the search for muonium-to-antimuonium conversion~\cite{Kawamura:2021lqk,Fukuyama:2021iyw,Heeck:2023iqc,Afik:2023vyl}.
The conversion process has not been studied by any experiment since 1999 and the latest limit on the conversion probability is $P_{\rm{M}\overline{\rm{M}}}<8.3\times10^{-11}$ (90\% C.L.)~\cite{Willmann:1998gd}.
Recently, we proposed a new muonium-to-antimuonium conversion experiment, MACE, aiming at improving the sensitivity to the spontaneously charged lepton flavor violating conversion process ($\rm{M}\rightarrow\overline{\rm{M}}$, or $\mu^+e^-\rightarrow\mu^-e^+$) by more than two orders of magnitude~\cite{Han:2021nod,Bai:2022sxq}.
MACE is designed to search for the rare process by distinguishing charge and kinematic properties of the decay products between muonium and antimuonium, where only vacuum muonium conversion events are of great interest.
These muonium-related experiments are benefited strongly from the increase of the muonium yield in vacuum.

How can we produce vacuum muonium?
A muon beam is injected into a specific material and a muon may spontaneously capture an electron to compose a muonium atom.
After that, some of produced muonium atoms would eventually escape from the material and emit into vacuum.
The target material is chosen to be porous and inactive such as silica powder or silica aerogel.
As a matter of fact, the silica powder was used in the latest experiment searching for muonium-antimuonium conversion~\cite{Willmann:1998gd}.
However, the silica powder is fragile and has relatively low vacuum muonium yield.
The silica aerogel shares the similar composition but takes a rich porous meso-structure, suitable for the muonium target.
Early in 1992, the muonium emission efficiency in a silica aerogel was measured for the first time and the result was comparable to that of the silica powder, even though a muon beam with higher momentum was used (which is likely to be adverse to muonium emissions into vacuum as discussed in \autoref{sec:optimization})~\cite{SCHWARZ1992244}.
After 30 years, a new measurement was carried out and the result demonstrated statistically that up to 3 muonium atoms could emit into vacuum out of 1000 stopped muons~\cite{Bakule:2013poa}.
Keep in mind that most of the muonium atoms formed in the material are still trapped~\cite{Beer:1986qr,Antognini:2011ei,Bakule:2013poa,Beare:2020gzr}.
There remains a plenty of room for improvement in the muonium emission into vacuum and a new approach is expected.

\begin{figure}[!t]
    \includegraphics[width=0.80\textwidth]{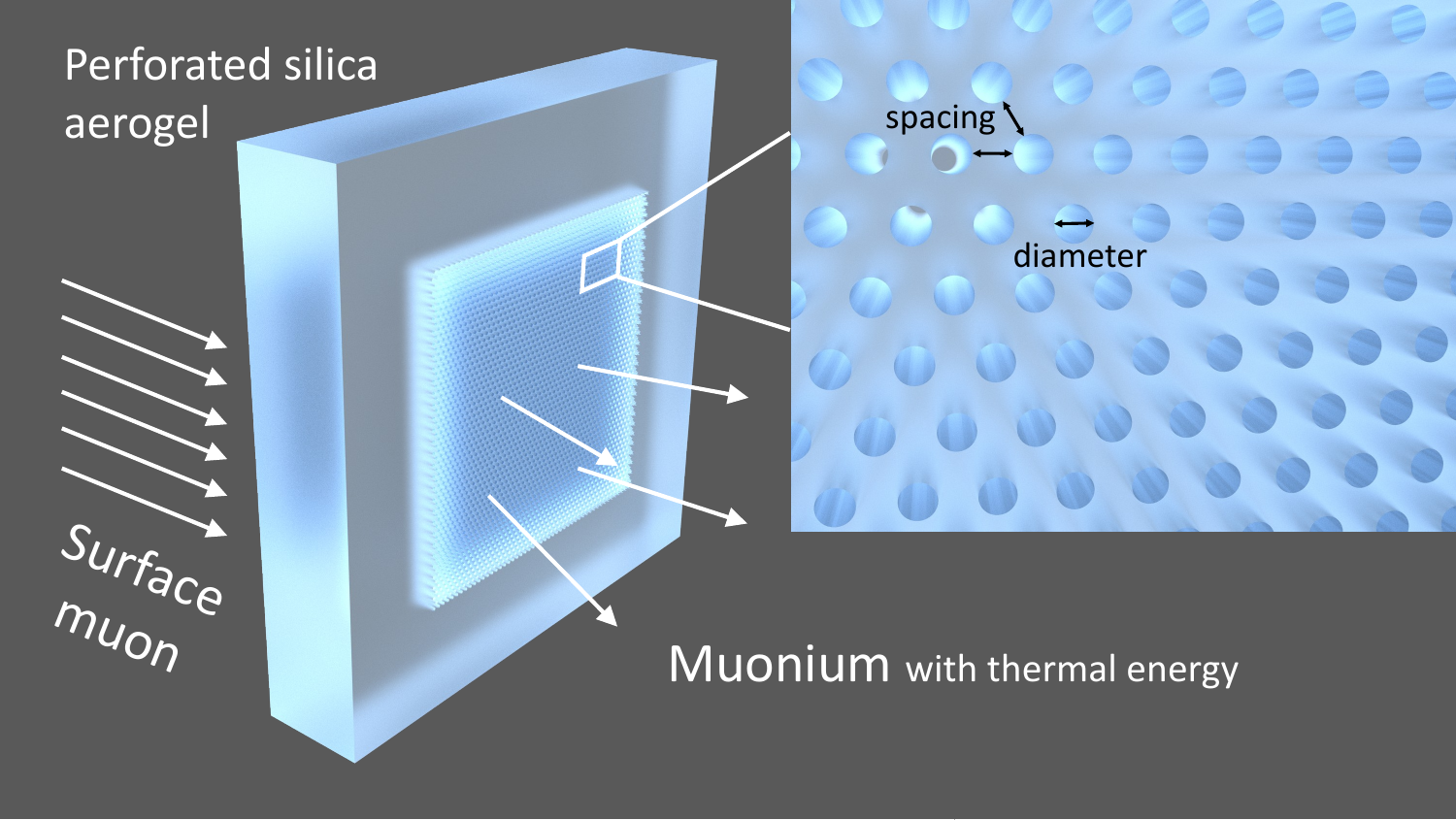}
    \caption{\label{fig:TargetSchemanticDiagram}The schematic diagram of muonium emission in the silica aerogel target.}
\end{figure}

G. A. Beer \textit{et al.} proposed a laser-ablation approach aimed at increasing the muonium emission efficiency.
In this approach, an aerogel target is ablated with a femtosecond pulsed laser so that the target surface at the beam downstream side is perforated (\cref{fig:TargetSchemanticDiagram})~\cite{Beer:2014ooa}.
Laser ablates nearly cylindrical holes with diameters ranging from tens to hundreds of micrometers and depths of a few millimeters.
The holes are arranged in an equilateral triangular lattice, perpendicular to the surface of the target with a spacing of tens to hundreds of micrometers.
The perforated structure significantly enhances the diffusion of muonium and dramatically increases its vacuum yield.
Multiple ablation structure parameters were tested in TRIUMF and an emission efficiency of up to 2\% was achieved, indicating an order of magnitude improvement~\cite{Beer:2014ooa}.
The authors also attempted to reveal the emission mechanism of muonium by proposing and analyzing the opening fraction, a characteristic parameter, to figure out the dependency of muonium emission efficiency.
However, the latest results roughly showed a positive correlation between the opening fraction and yield while more complex dependencies remained blank.
The insufficient number of samples also led to an incomplete understanding of the underlying mechanism.
In addition to the experiments conducted at TRIUMF, measurements on muonium emission efficiency in perforated aerogel were also conducted in PSI~\cite{Antognini:2022ysl}.
In the PSI experiment, a slow muon beam ($p_{\mu^+}=11$--13~\text{MeV}/$c$) entered the front surface of the aerogel sample, and muonium atoms were emitted into vacuum upstream.
However, the emission efficiency depends on the emission model parameters and further discussion on the enhancement of muonium emission by perforated structures is missing.
We also note that an independent simulation study was carried out to model the muonium diffusion in silica aerogel and was applied to a novel multi-layer target design~\cite{Zhang:2022ilj}.
Their study modelled the transport of muonium as a diffusion process with a thermal momentum distribution.
However, they incorporated the effect from enlarged diffusion constants while detailed considerations of perforated structures remained blank.
In this study, we will model muonium diffusion as a three-dimensional off-lattice thermal random walk and propose a new muonium tracking algorithm that comprehensively considers the perforated structure.
We will simulate the muonium diffusion inside perforated silica aerogel with the measured mean free path rather than fitting the diffusion constant so that we will carry out the first explicit dependency of the muonium yield on the perforation parameters.
Furthermore, the simulation method will be validated and applied to a muonium yield optimization.
We will attempt to analyze and understand the dependencies and the underlying mechanisms involved in the muonium emission.
This study will further guide the development and optimization of the muonium target, and pave the way for the design and optimization of MACE.
Our method will not only benefit the precision measurement and the search for new physics in the muon sector, and the development on cutting-edge muon spin relaxation/rotation/resonance ($\mu$SR) techniques~\cite{Woodle:1988Msr,10.1063/5.0004210,Pant:2021lch,Hillier:2022musr,Ni:2023ppw}, but also be meaningful to the positronium physics and technologies, including the positronium production and emission~\cite{Cassidy:2010cea,Ferragut:2011vrl,Girardi-Schappo:2013elo,Deller:2015txa}, positronium spectroscopy and annihilation~\cite{Gurung:2020hms,Moskal:2021kxe,Adkins:2022omi,Sheldon:2023vgh}, antihydrogen gravity experiment (AE$\bar{\text{g}}$IS)~\cite{Ferragut:2011vrl,AEGIS:2012eto,Volponi:2023xpz}, and positron emission tomography (PET)~\cite{Bass:2023dmv}.

This article is organized as follows: In \autoref{sec:SimulationMethod}, the theoretical formalism and the simulation method are described. In \autoref{sec:SimulationValidation}, we present the validation of the simulation. We then apply the new algorithm to optimize the vacuum muonium yield in \autoref{sec:optimization}. The findings of this work are finally summarized in \autoref{sec:summary}.

\section{Muonium diffusion and its simulation}
\label{sec:SimulationMethod}

Silica aerogel is an inactive porous material with a rich meso-structure.
It is a matter similar to common gels, but liquid component are replaced by gas, which makes its porosity extremely high (80\%--99.8\%) but density extremely low ($\mathcal{O}$(mg/cm$^3$))~\cite{SOLEIMANIDORCHEH200810,Akhter2021,PATIL20212981}.
Microscope photos demonstrate that a silica aerogel has a complex network structure, with a characteristic scale of tens to hundreds of nanometers.
The silica network structure allows molecules or atoms diffuse inside the inter-granular void~\cite{HOSTICKA1998293}.

If the aerogel is placed in vacuum for a sufficient period, air molecules inside would diffuse and come out of the material.
The aerogel would become a ``vacuumgel'' and muonium can move freely inside the inter-granular void until colliding with silica.
After a collision, muonium will mostly be inelastically scattered instead of being captured.
This proposition is supported by density functional theory (DFT) calculations of hydrogen atom dynamics inside silica.
Based on DFT results, we can infer that a majority of muonium atoms can disassociate from silica and remain its state neutral~\cite{PhysRevB.92.014107}.
Independent experimental results on absorption of hydrogen by silica aerogel also demonstrate that significant absorption of hydrogen only takes place at a very low temperature, supporting this proposition to a certain extent~\cite{10.1063/1.5020910}.
Therefore, muonium atoms experience a cycle of free motion -- scattering -- free motion process in a silica aerogel before decays and it can be modelled as a three-dimensional off-lattice random walk.

Following this principle, a recursive equation of motion describing a muonium atom inside a silica aerogel can be written as
\begin{equation}\label{eq:MuoniumRandomWalkProcess}
    \begin{aligned}
        \Delta t_i           & =\frac{r_i}{v_i}~,\\
        t_{i+1}              & =t_i+\Delta t_i~,\\
        \boldsymbol{x}_{i+1} & =\boldsymbol{x}_i+\boldsymbol{v}_i\Delta t_i~, \\
    \end{aligned}
\end{equation}
where $\left(t_i,\boldsymbol{x}_i\right)$, $r_i$, $\boldsymbol{v}_i$ and $v_i=\lVert\boldsymbol{v}_i\rVert$ are the space-time coordinates, the free path, the velocity and the velocity magnitude at the $i$-th step, respectively.
Both $r_i$ and $\boldsymbol{v}_i$ are random variables sampled from the corresponding distribution at each step.
Considering the randomly connected network structure in a silica aerogel, it is convinced that the free path follows an exponential distribution with a parameter $\lambda$
\begin{equation}\label{eq:MuoniumFreePathDistribution}
    f_r(r)=\frac{1}{\lambda}\exp{\left(-\frac{r}{\lambda}\right)}~,
\end{equation}
where $\lambda=\langle r\rangle$ is the mean free path depending on the structure of a specific aerogel.
The magnitude of $\lambda$ is positively correlated with the size of meso-cavities inside the aerogel.

\begin{figure}
    \centering
    \includegraphics[width=0.9\textwidth]{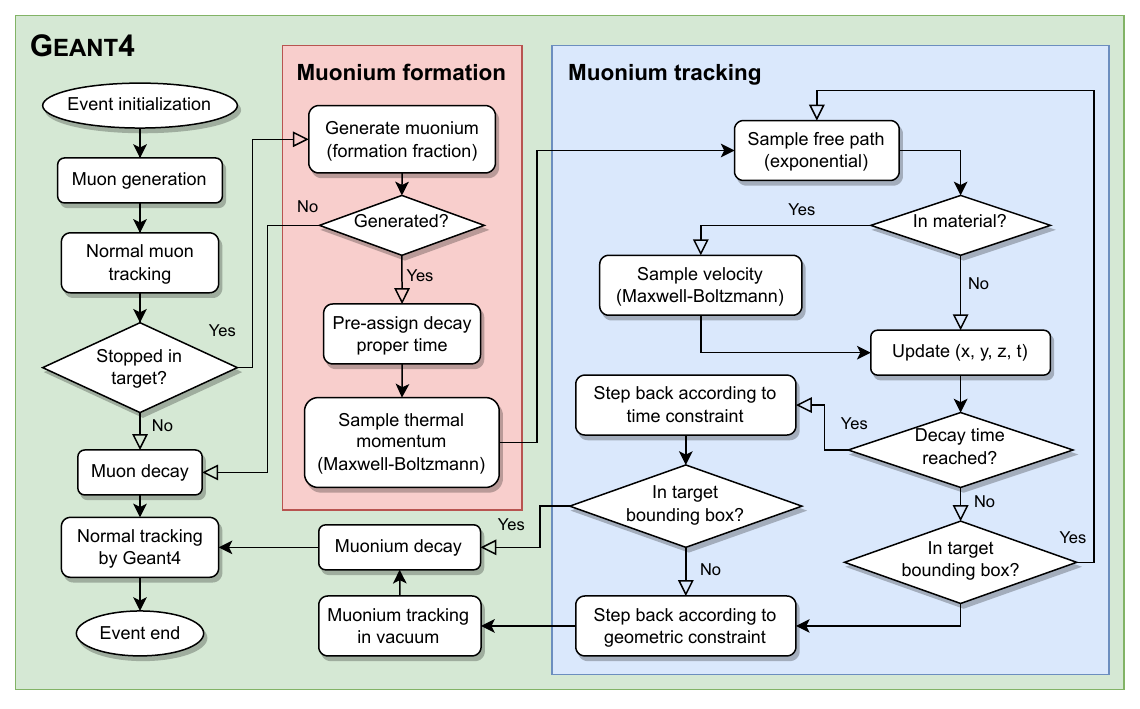}
    \caption{\label{fig:SimTarget}The muonium formation and tracking algorithm in the simulation.}
\end{figure}

It comes to the velocity distribution now.
Inside an aerogel, muonium atoms experience a series of inelastic scattering on silica structures, and the emission velocity distribution after scattering is related to the interaction between the muonium atoms and silica.~\cite{ZENG1995264,PhysRevB.92.014107,YUE20185}.
Therefore, in principle, the exact distribution should be determined by experimental data or \textit{ab inito} calculations.
As an approximation, it is believed that the ensemble of muonium atoms and target molecules quickly reaches the thermal equilibrium~\cite{ZENG1995264}.
Therefore, the kinetic energy of muonium atoms should follow the form of the Boltzmann distribution, and the velocity distribution takes the form
\begin{equation}\label{eq:MuoniumVelocityDistribution}
    f_{\boldsymbol{v}}(\boldsymbol{v})=\left(\frac{m_\text{M}}{2\pi k_\text{B} T_\text{eff}}\right)^{3/2}\exp{\left(-\frac{m_\text{M}\boldsymbol{v}^2}{2 k_\text{B} T_\text{eff}}\right)}~,
\end{equation}
where $m_\text{M}$ is the mass of muonium, $k_\text{B}$ is the Boltzmann constant and $T_\text{eff}$ is the effective temperature.
It is known that the room temperature will overestimate the muonium decay time.
One of the interpretation is the underestimation of kinetic energy.
We can always correct the effect by elevating the temperature~\cite{Bakule:2013poa,Zhang:2022ilj}.
The experimental work of Antognini \textit{et al.} found the temperature to be 400~K, by considering a simplified emission model with a given diffusion time $\Delta t_\text{diff}=200~\text{ns}$~\cite{Antognini:2022ysl}.
However, their emission model has correlations between $\Delta t_\text{diff}$ and temperature.
Therefore, the temperature depends on the choice of $\Delta t_\text{diff}$.
Another simulation study presented by Ce Zhang \textit{et al.} estimated the temperature to be around 320~K ($320\pm2~\text{K}$ for flat target and $322\pm2~\text{K}$ for perforated target) with datasets measured at TRIUMF.
In the model, the diffusion time is naturally included and the temperature is determined by data fitting.
In out study, we choose $T_\text{eff}=322~\text{K}$ to perform the following simulations.
Equation \eqref{eq:MuoniumRandomWalkProcess}--\eqref{eq:MuoniumVelocityDistribution}, as an iterative scheme of the special off-lattice random walk, simulates the motion of muonium until it reaches the material surface or decays.

To perform accurate simulations in a perforated structure, it is also important to track muonium outside the material reliably.
The commonly used method for navigating and tracking particles in a complex geometry is to calculate the geometric relations between particles and the voxelized geometry (e.g. \GeantFour~\cite{GEANT4:2002zbu,Allison:2006ve,Allison:2016lfl}).
However, this method can hardly handle a perforated target with tens of thousands of micro holes, since neither the full geometry can be easily constructed nor an efficient navigation or tracking can be performed.
Therefore, an alternative method is needed.
Our solution is to use a combinatorial tracking method.
The voxelized-geometry-based tracking method controls muonium motion outside the target bounding box while a Boolean-expression-based tracking method controls muonium motion inside.
This combination can resolve the realizability and performance issues raised in the voxelized-geometry-based tracking method.

To implement the method, the first question is how to describe the target full geometry in the combinatorial tracking system.
The solution involves a Boolean expression to describe the perforated structure and a bounding box (a \GeantFour volume) to describe the target region where the Boolean expression should be enabled.
The Boolean expression divides the target volume into two different sub-regions: a \texttt{true} region representing the material where muonium atoms follow the random walk, and a \texttt{false} region representing the vacuum where muonium atoms move freely.
During each step of the random walk, the Boolean expression is calculated to determine the inclusion of the muonium in the material and the state of motion.
If a muonium atom appears inside the material, a free path and a velocity are sampled and the spacetime coordinates are updated.
Otherwise, only a free path is sampled and the state of motion is updated with the current velocity.
The process will loop until any of the termination conditions, i.e., the muonium decays or escapes from the target bounding box, is triggered.
Then the Boolean-expression-based tracking comes to an end and the tracking procedure is returned to the voxelized-geometry-based method (\GeantFour).
The muonium formation and tracking algorithm in the simulation is shown in \cref{fig:SimTarget}.

This combinatorial tracking method combines the power of \GeantFour and the Boolean-expression-based algorithm in the same procedure, allowing us to successively simulate a muon track and a muonium track.
We are able to simulate the dependency of the muonium yield with regards to different muon beam conditions conveniently and precisely, so that the approach can be validated with experimental data.
The large-scale simulation can be performed in a local supercomputer following the software deployment strategy in Ref.~\cite{Chen:2023qcp}.

\section{\label{sec:SimulationValidation}Validation of simulation scheme}

After a discussion of how to simulate the muonium diffusion in a perforated target, this section demonstrates a validation of the simulation method.
The previous experimental studies, utilizing aerogel targets with many tiny blind holes arranged in an equilateral triangular lattice at the downstream surface, have accumulated vacuum muonium production data in different experimental setups.
Two datasets can be used for the validation: one is the vacuum muonium yield data~\cite{Beare:2020gzr}, and the other is the spacetime distribution of muonium decays~\cite{Bakule:2013poa,Beer:2014ooa}.
Simulations will be performed and corresponding datasets will be derived by analyzing the simulation data.

The model described in \autoref{sec:SimulationMethod} contains two free parameters: a mean free path $\lambda$ and a temperature $T$.
While the temperature is chosen to be 322~K according to the previous fit~\cite{Zhang:2022ilj}, the mean free path is estimated by the empirical scaling formula $\lambda = \lambda_0\left(\rho_0/\rho\right)^{1.5}$~\cite{Bakule:2013poa}.
The reference mean free path $\lambda_0$ and the reference target density $\rho_0$ are 226~nm and $29~\text{mg}/\text{cm}^3$, respectively.
Since the target densities involved in this validation are close to the reference values, the scaling formula can be applied.
The perforated structure is simplified as cylindrical blind holes arranged in an equilateral triangular pattern, with its axis perpendicular to the target surface, as shown in \cref{fig:TargetSchemanticDiagram}.
This structure is represented completely by a Boolean expression and a target bounding box (\GeantFour \texttt{G4Box}).
The spacing, diameter, and depth of the holes are geometric parameters kept the same as the experiment.
The remaining simulation settings are also kept the same as the corresponding benchmark experiment.

\begin{figure}
    \centering
    \subfloat[Flat target, region 1]{
        \includegraphics[width=0.31\textwidth]{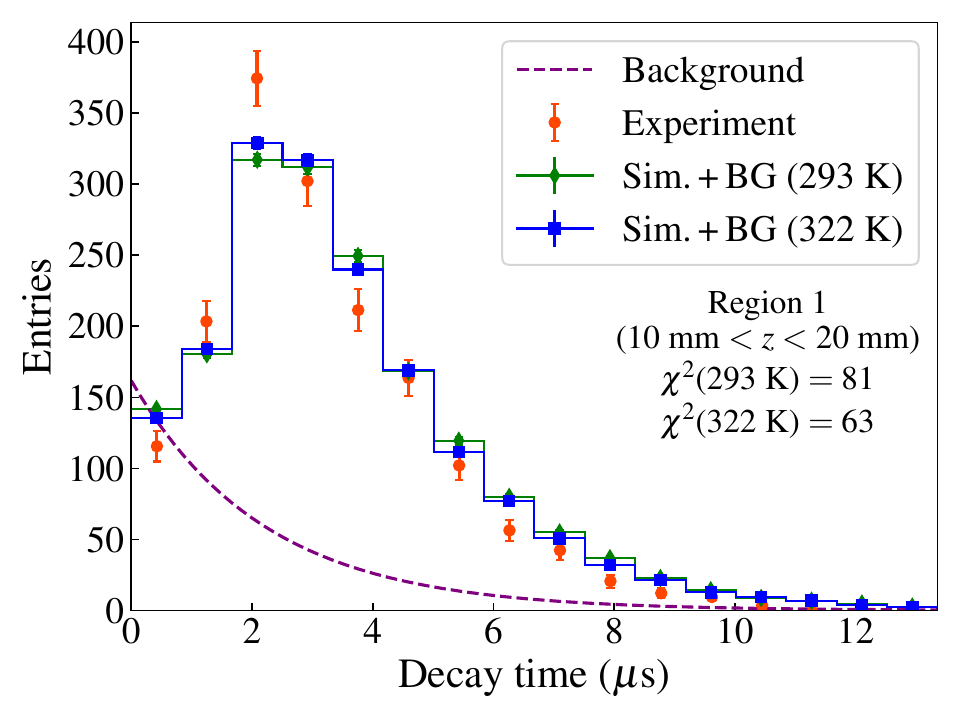}
    }
    \subfloat[Flat target, region 2]{
        \includegraphics[width=0.31\textwidth]{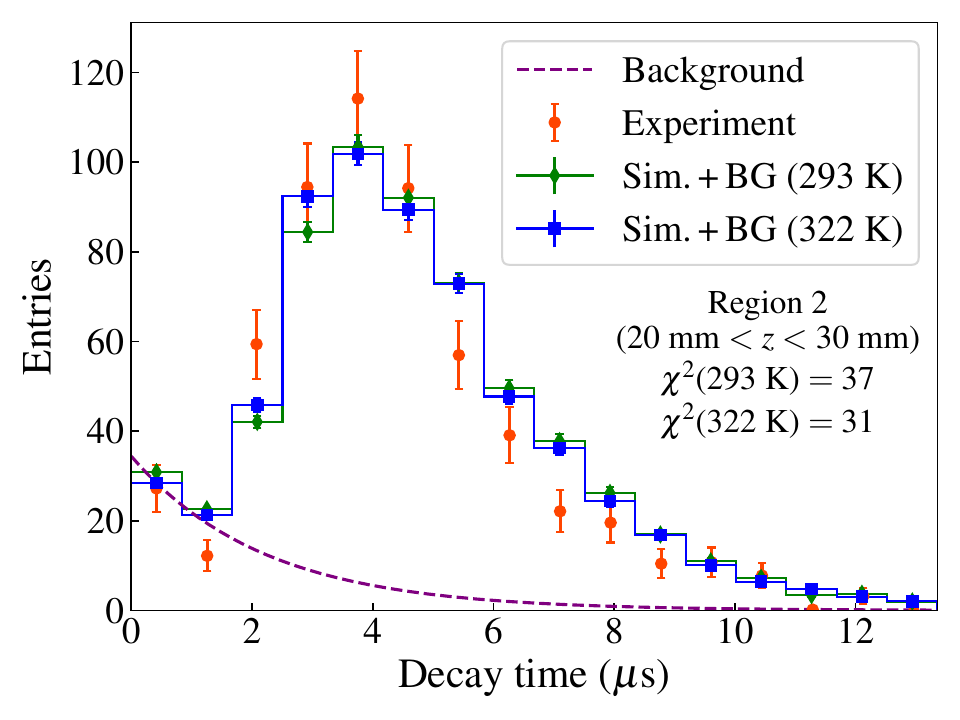}
    }
    \subfloat[Flat target, region 3]{
        \includegraphics[width=0.31\textwidth]{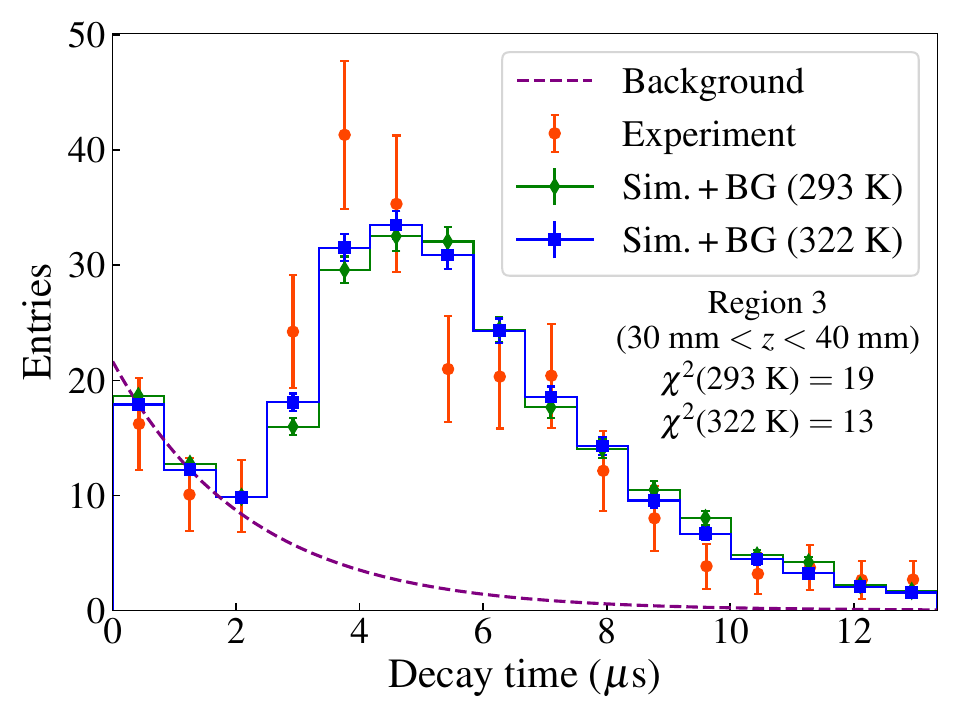}
    }\\
    \subfloat[Perforated target, region 1]{
        \includegraphics[width=0.31\textwidth]{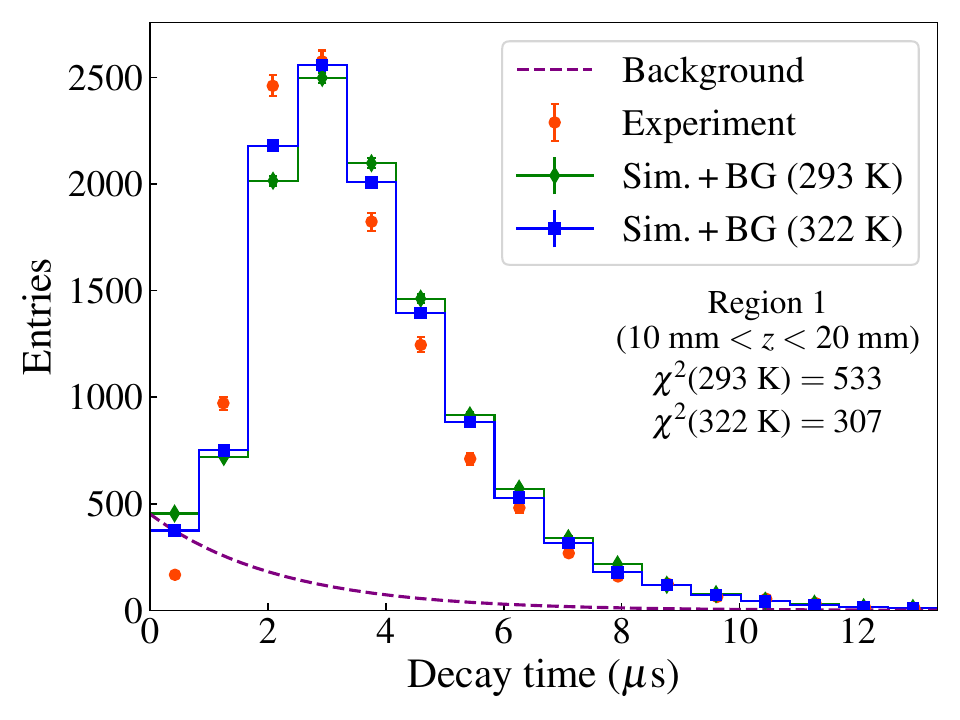}
    }
    \subfloat[Perforated target, region 2]{
        \includegraphics[width=0.31\textwidth]{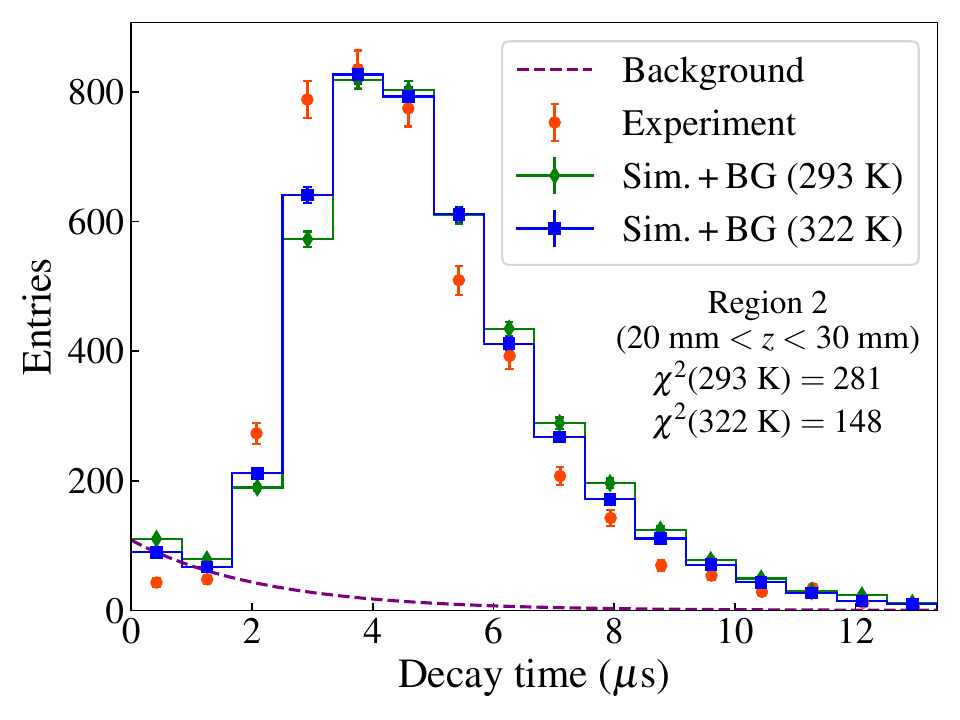}
    }
    \subfloat[Perforated target, region 3]{
        \includegraphics[width=0.31\textwidth]{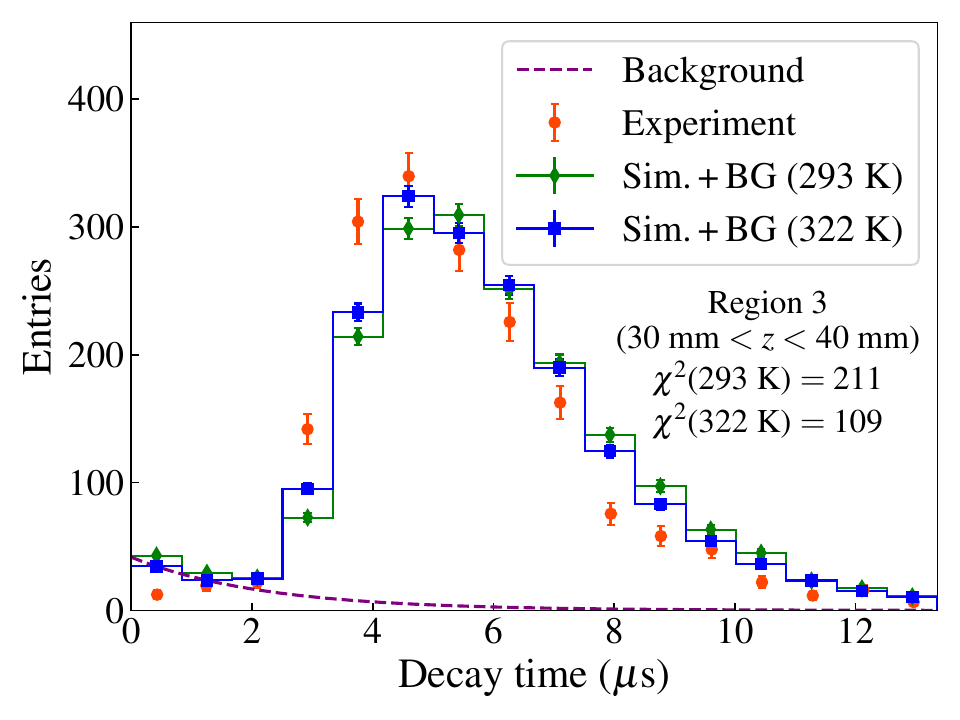}
    }
    \caption{\label{fig:SpacetimeValidation}The spacetime distribution of muonium decay. The simulation histogram and an exponential muon decay background are fit to the experimental histogram.}
\end{figure}

We first validate the spacetime distribution of muonium decay by comparing it with measurements in TRIUMF~\cite{Bakule:2013poa,Beer:2014ooa}.
In the simulation, a $\sigma_{xy}=5~\mathrm{mm}$ Gaussian profiled collimated surface muon beam carrying 22.9~MeV/$c$ momentum with 1.5\% spread (rms) is incident on the target.
The target thickness is 6.9~mm and perforated by 270~$\mathrm{\mu m}$ diameter and 30~$\mathrm{\mu m}$ spacing holes, from which some of the muonium atoms emit into the surrounding vacuum and decay.
These decays are divided into three different space regions ranging from $z=10~\mathrm{mm}$ to 40~mm, as shown in the simulation results (\cref{fig:SpacetimeValidation}).
The simulated distribution and an exponential muon decay background are fitted to the experimental histogram.
A room temperature (293~K) and an elevated temperature (322~K, according to Zhang's fit~\cite{Zhang:2022ilj}) are used in the simulation.
The result shows agreement between simulation and experiment with $\chi^2(322~\mathrm{K})<\chi^2(293~\mathrm{K})$ for all cases, as we mentioned above.
We believe the goodness of fit can be further improved by tuning temperature for each case exclusively.
In this validation case, we deal with the decay spacetime distribution of muonium emission.
Our primary goal is to simulate the vacuum yield of the perforated target.
One might ask whether the small difference in the decay spacetime distribution will significantly affect the yield result.
We will answer this question by a second round of validation on the vacuum muonium emission efficiency.

\begin{figure}
    \centering
    \includegraphics[width=0.7\textwidth]{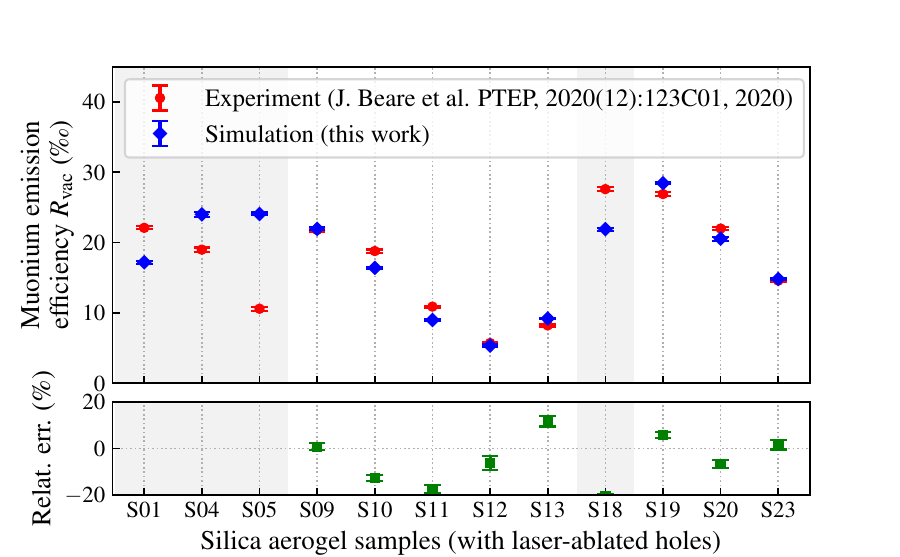}
    \caption{\label{fig:YieldValidation} A comparison of muonium emission efficiency in simulation and experiment.}
\end{figure}

We now simulate vacuum muonium emission efficiency of perforated targets with different ablation schemes (S01, S04, S05, S09--S13, S18--S20, and S23 in Ref.~\cite{Beare:2020gzr}) and compare them with measurement results in TRIUMF~\cite{Beare:2020gzr}.
The simulation setup follows the experiment.
In the simulation, a collimated surface muon beam with 23~MeV/$c$ momentum and 2\% momentum spread (rms) hits the target with a Gaussian profile of $\sigma_{xy}=5~\mathrm{mm}$.
A plastic scintillator is placed at $z=-12~\mathrm{mm}$, with 40~mm width and 380~$\mu$m thickness.
The target geometry is identical to the experiment and its downstream surface is placed at $z=0$.
Muonium decays at $z>0$ are counted, and are used to calculate muonium emission efficiency by a ratio of the number of muonium decays in vacuum over the number of decays in target.
The simulation results are presented in \cref{fig:YieldValidation}, and errors between the simulation and the experiment are mostly within 20\%, indicating a certain agreement.
The significant deviations in the samples (S01, S04, S05, and S18) are currently not well explained.
One of the possible reason might be related the deformation of the real target samples, which can hardly covered in the simulation.
We tend to consider the deviations as the systematic errors, which will not affect the reliability of the simulation.
In conclusion, the straightforward simulations now can describe the transport of muonium well in complex aerogel geometry and provide reasonable yield estimations.
Precision measurement experiments might need calibrated simulations in details.

\section{Simulation-guided yield optimization}

\label{sec:optimization}
\begin{figure}[!t]
    \centering
    \subfloat[Total yield $Y_\text{tot}$.]{
        \includegraphics[width=0.3\textwidth]{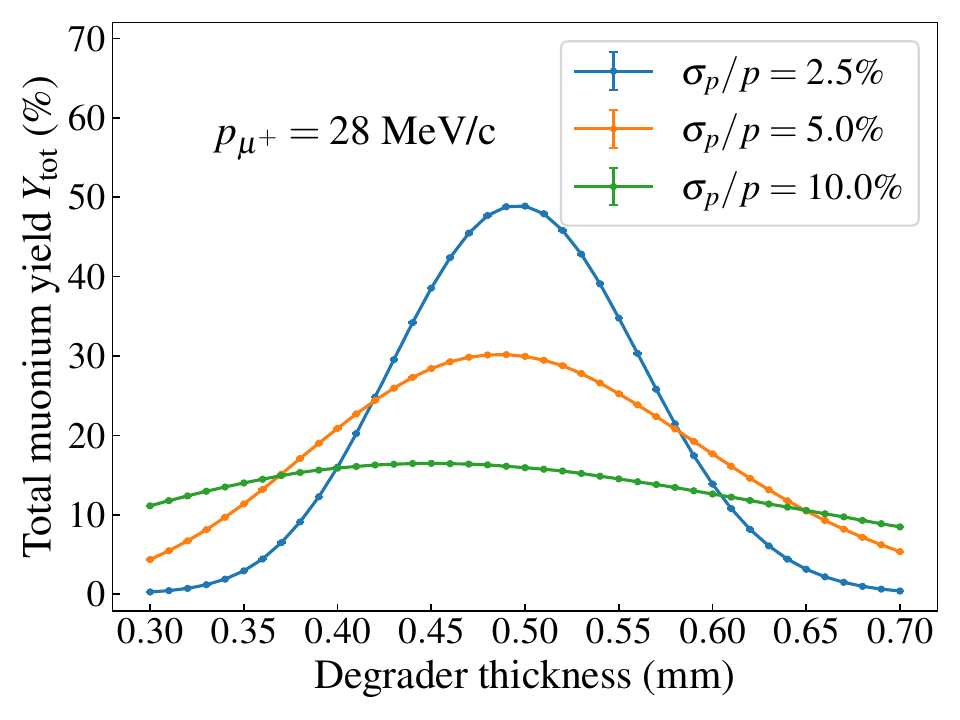}
    }
    \subfloat[Emission efficiency $R_\text{vac}$.]{
        \includegraphics[width=0.3\textwidth]{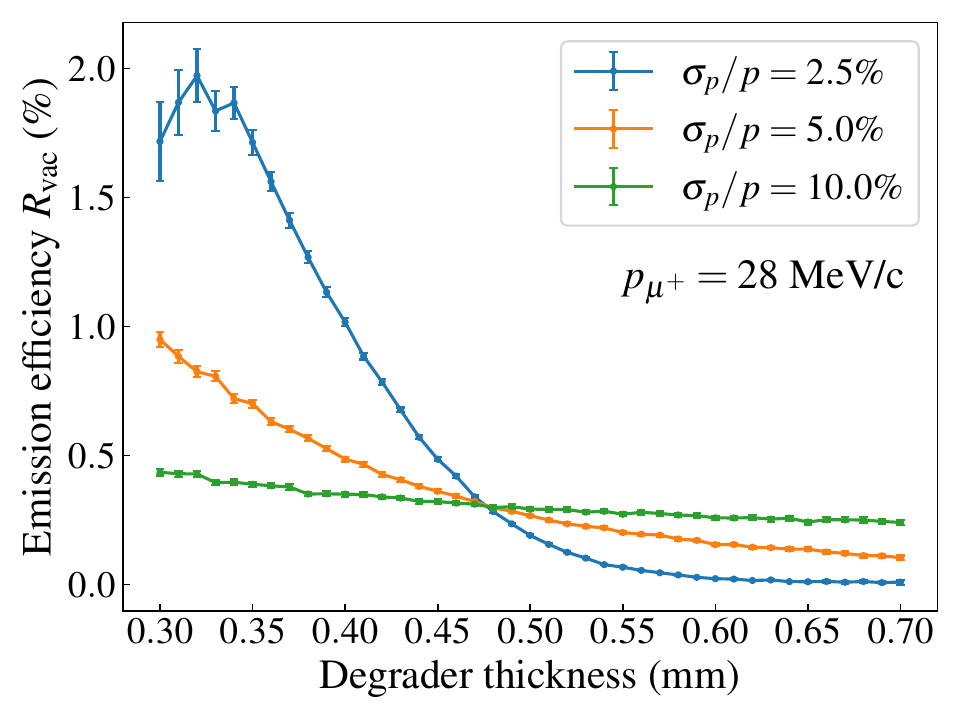}
    }
    \subfloat[Vacuum yield $Y_\text{vac}$.]{
        \includegraphics[width=0.3\textwidth]{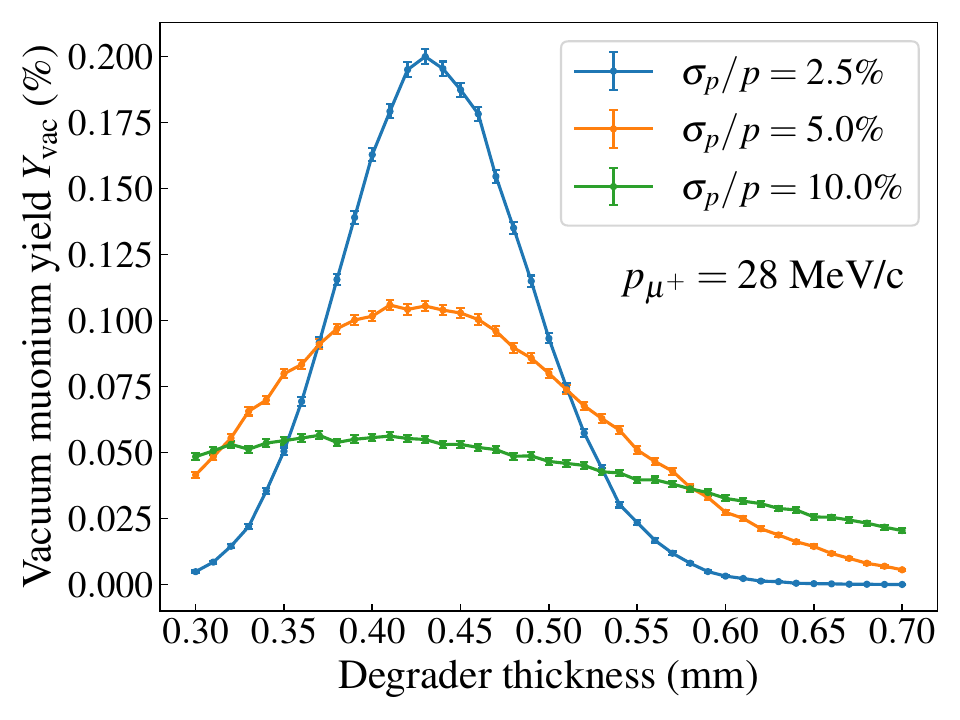}
    }
    \caption{\label{fig:al_degrader_thickness}Muonium production and emission from a flat silica aerogel target with different aluminium degraders ($t_\text{target}=10~\text{mm}, \rho_\text{target}=30~\text{mg}/\text{cm}^3, \lambda=200~\text{nm}$, and $T=322~\text{K}$).}
\end{figure}

\begin{figure}
    \centering
    \subfloat[Muonium total yield $Y_\text{tot}$ ($\sigma_p/p=2.5\%,\ d=1~\mathrm{mm}$)]{
        \includegraphics[width=0.3\textwidth]{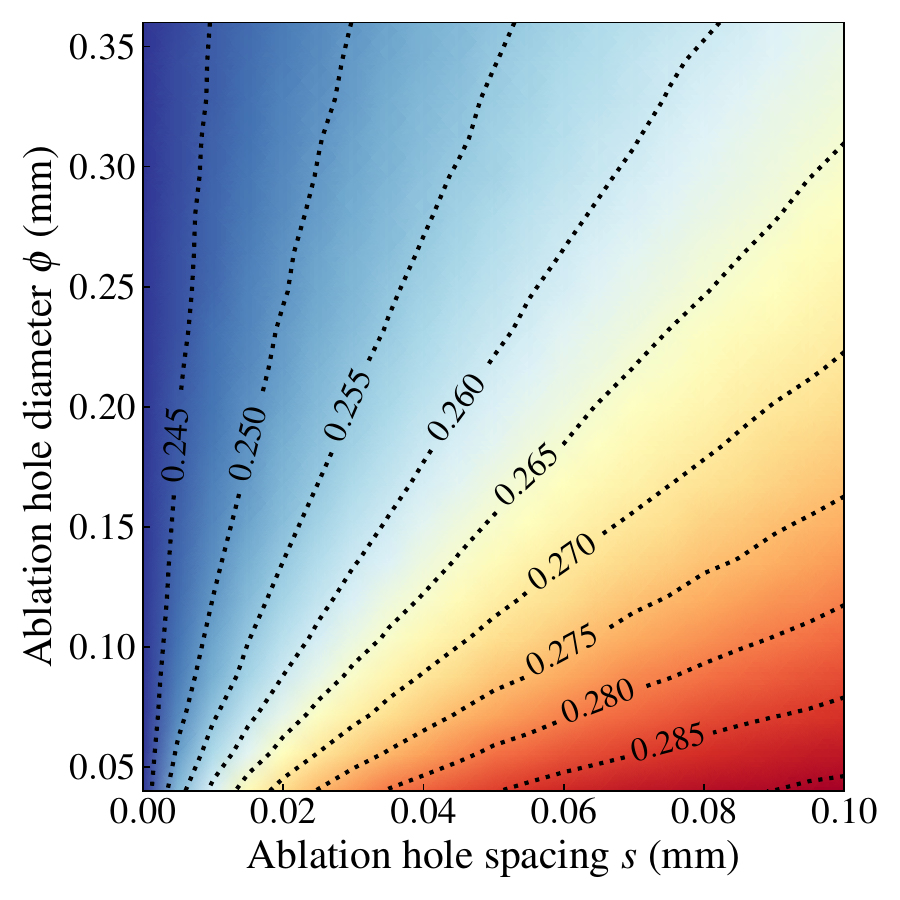}
    }
    \subfloat[Emission efficiency $R_\text{vac}$ ($\sigma_p/p=2.5\%,\ d=1~\mathrm{mm}$)]{
        \includegraphics[width=0.3\textwidth]{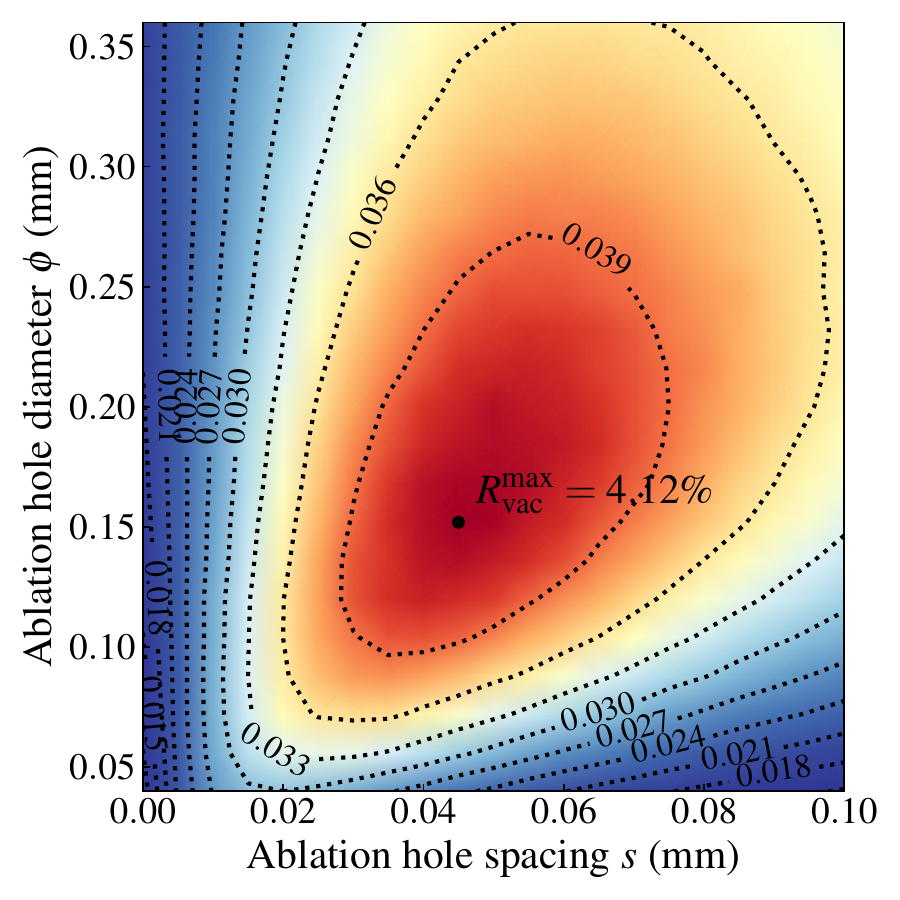}
    }
    \subfloat[Vacuum muonium yield $Y_\text{vac}$ ($\sigma_p/p=2.5\%,\ d=1~\mathrm{mm}$)]{
        \includegraphics[width=0.3\textwidth]{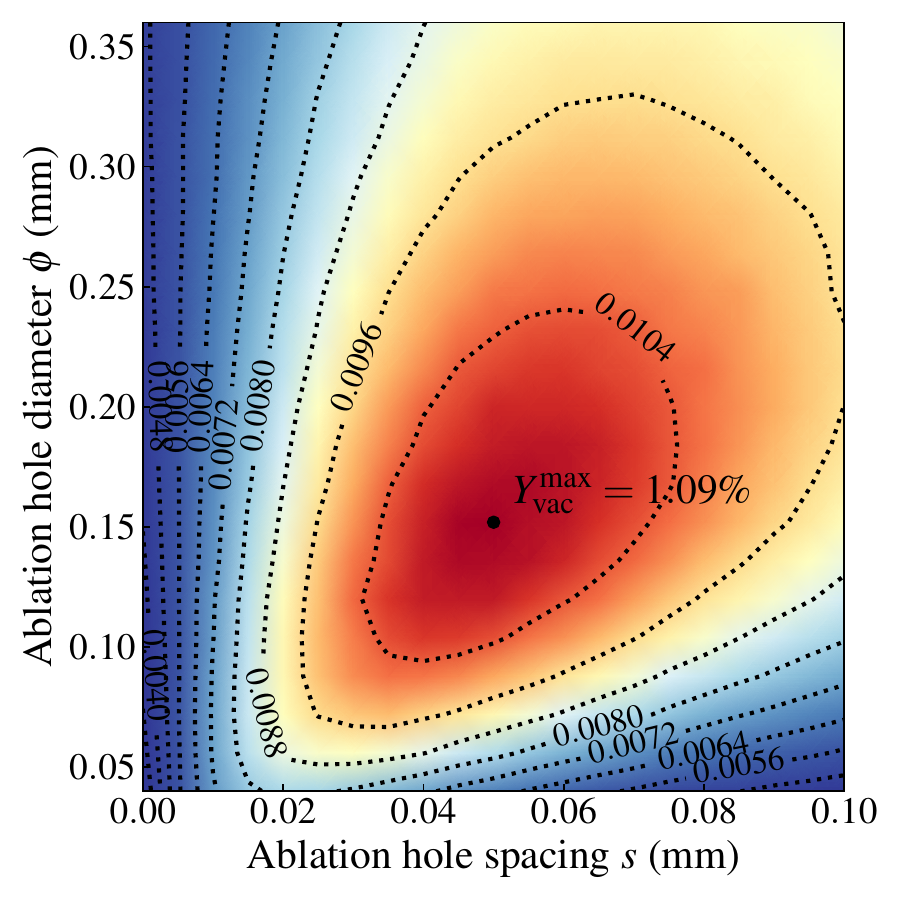}
    }\\
    \subfloat[Muonium total yield $Y_\text{tot}$ ($\sigma_p/p=10\%,\ d=5~\mathrm{mm}$)]{
        \includegraphics[width=0.3\textwidth]{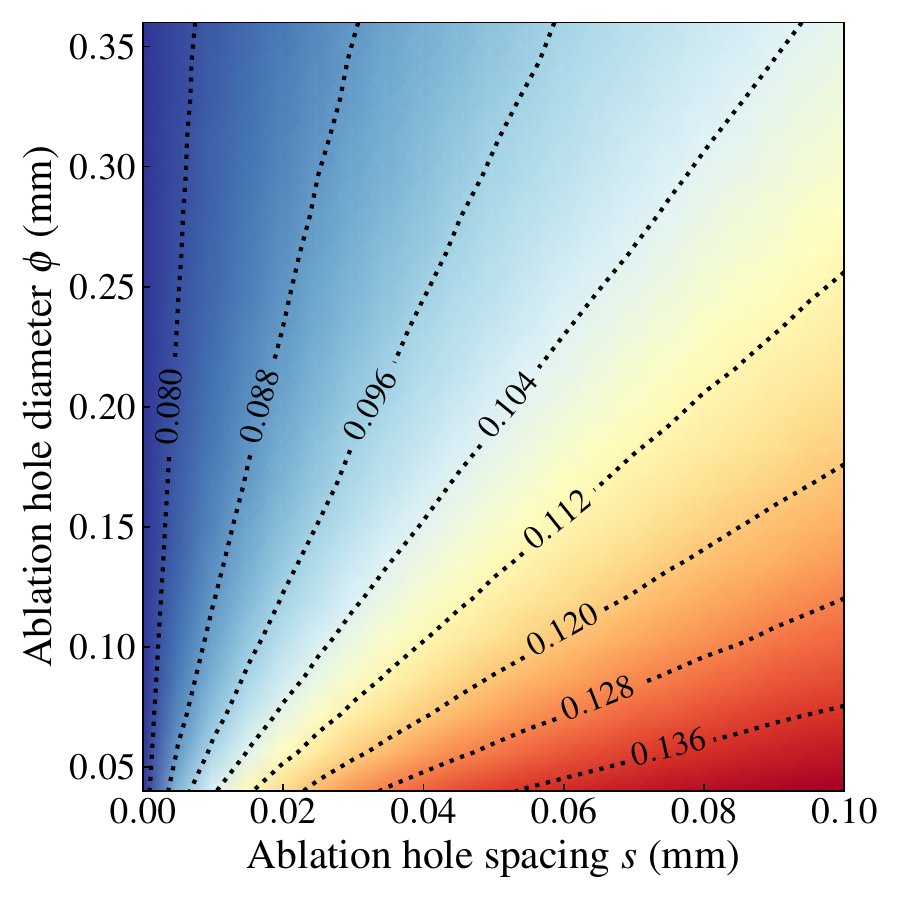}
    }
    \subfloat[Emission efficiency $R_\text{vac}$ ($\sigma_p/p=10\%,\ d=5~\mathrm{mm}$)]{
        \includegraphics[width=0.3\textwidth]{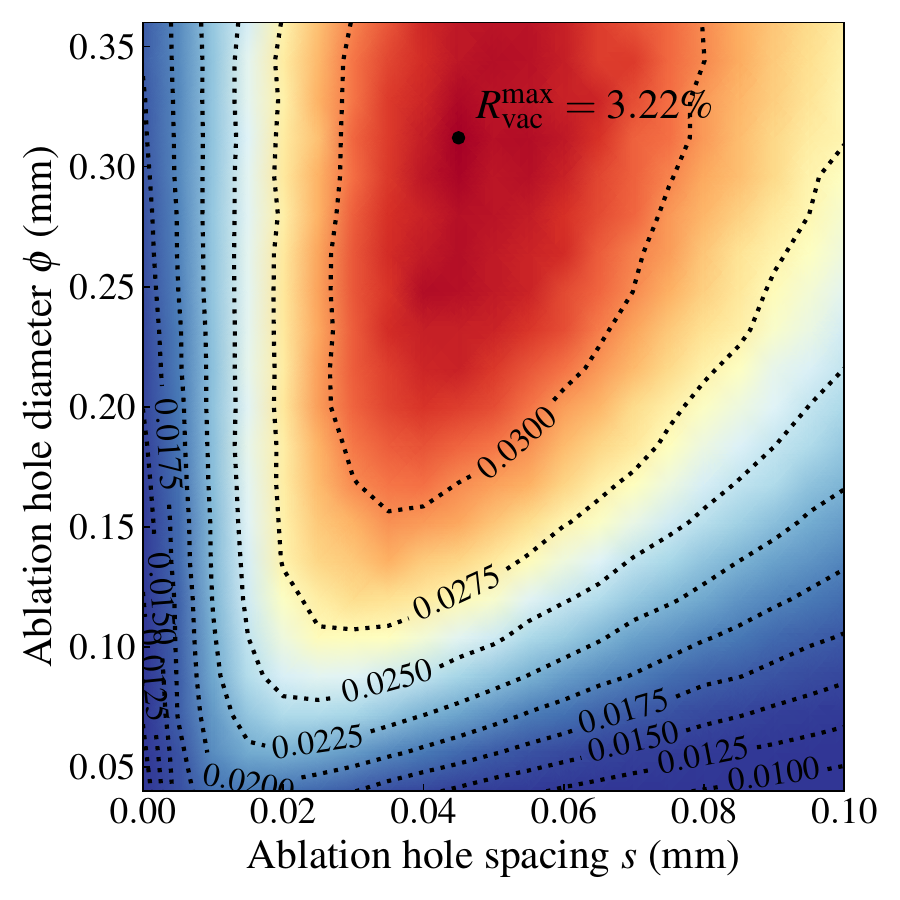}
    }
    \subfloat[Vacuum muonium yield $Y_\text{vac}$ ($\sigma_p/p=10\%,\ d=5~\mathrm{mm}$)]{
        \includegraphics[width=0.3\textwidth]{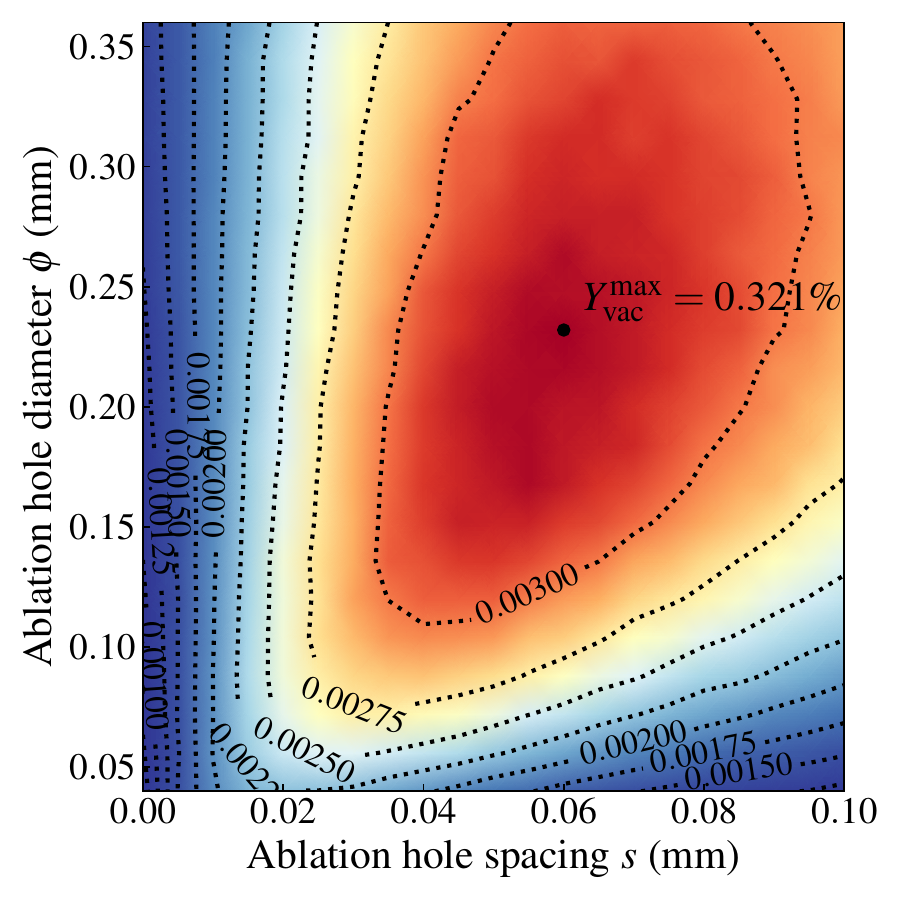}
    }
    \caption{\label{fig:scan_simulation}Projection of muonium total yield, muonium emission efficiency, and vacuum muonium yield under different beam conditions and target geometries.}
\end{figure}

\begin{table}[!t]
    \centering
    \caption{\label{tab:maxRvac}Simulation of maximum muonium emission efficiency and corresponding optimal spacing and diameter with different beam condition.}
    \begin{tabular}{c|c|c|c|c|c|c}
        \hline\hline
        \multirow{2}*{Data}                              & $p_\text{beam}$ & $\sigma_{p_\text{beam}}/p_\text{beam}$ & Depth           & Spacing      & Diameter        & Max emission                                     \\
                                                         & (MeV/$c$)       & (\%)                                   & $d$ (mm)        & $s$ ($\mu$m) & $\phi$ ($\mu$m) & efficiency $R_\text{vac}$ (\%)                   \\
        \hline\hline
                                                         &                 &                                        & 1               & $45 \pm 5$   & $152 \pm 16$    & $4.124 \pm 0.007$                                \\
                                                         &                 & 2.5                                    & 2               & $50 \pm 5$   & $232 \pm 16$    & $4.980 \pm 0.009$                                \\
                                                         &                 &                                        & 5               & $35 \pm 5$   & $360 \pm 16$    & $7.92 \pm 0.02$                                  \\
        \cline{3-7}
                                                         &                 &                                        & 1               & $50 \pm 5$   & $168 \pm 16$    & $2.762 \pm 0.007$                                \\
        Simulation~\footnotemark[1]                      & 28              & 5                                      & 2               & $50 \pm 5$   & $248 \pm 16$    & $3.172 \pm 0.008$                                \\
                                                         &                 &                                        & 5               & $40 \pm 5$   & $312 \pm 16$    & $4.36 \pm 0.01$                                  \\
        \cline{3-7}
                                                         &                 &                                        & 1               & $50 \pm 5$   & $152 \pm 16$    & $2.166 \pm 0.007$                                \\
                                                         &                 & 10                                     & 2               & $50 \pm 5$   & $232 \pm 16$    & $2.451 \pm 0.008$                                \\
                                                         &                 &                                        & 5               & $45 \pm 5$   & $312 \pm 16$    & $3.22 \pm 0.01$                                  \\
        \hline
        Ref.~\cite{Beare:2020gzr} (S18)~\footnotemark[1] & 23              & 2                                      & 1               & $85$         & $165$           & $2.76 \pm 0.02$~\footnotemark[2]                 \\
        \hline
        Ref.~\cite{Beer:2014ooa}~\footnotemark[1]        & 28              & 5                                      & $4.75 \pm 0.25$ & $30$         & $270$           & $3.05 \pm 0.03$~\footnotemark[2]                 \\
        \hline
        Ref.~\cite{Antognini:2022ysl} (Aerogel-1)         & 12.5            & 3.4                                    & $4.5 \pm 0.5$   & $45 \pm 5$   & $105 \pm 5$     & $6.72 \pm 0.05^{+1.06}_{-0.76}$~\footnotemark[3] \\
        \hline\hline
    \end{tabular}
    \footnotetext[1]{Only statistical errors are shown.}
    \footnotetext[2]{Includes muonium decays within $10~\text{mm}<z<40~\text{mm}$ only.}
    \footnotetext[3]{Model-dependent assumptions of the temperature at 400~K and the diffusion time at 200~ns.}
\end{table}

\begin{table}[!t]
    \centering
    \caption{\label{tab:maxYvac}Simulation of maximum vacuum muonium yield and corresponding optimal spacing and diameter with different beam condition.}
    \begin{tabular}{c|c|c|c|c|c|c}
        \hline\hline
        \multirow{2}*{Data}                                       & $p_\text{beam}$ & $\sigma_{p_\text{beam}}/p_\text{beam}$ & Depth           & Spacing      & Diameter        & Max vacuum                                         \\
                                                                  & (MeV/$c$)       & (\%)                                   & $d$ (mm)        & $s$ ($\mu$m) & $\phi$ ($\mu$m) & yield $Y_\text{vac}$ (\%)                          \\
        \hline\hline
                                                                  &                 &                                        & 1               & $50 \pm 5$   & $152 \pm 16$    & $1.092 \pm 0.002$                                  \\
                                                                  &                 & 2.5                                    & 2               & $55 \pm 5$   & $184 \pm 16$    & $1.134 \pm 0.002$                                  \\
                                                                  &                 &                                        & 5               & $55 \pm 5$   & $184 \pm 16$    & $1.122 \pm 0.002$                                  \\
        \cline{3-7}
                                                                  &                 &                                        & 1               & $50 \pm 5$   & $152 \pm 16$    & $0.583 \pm 0.001$                                  \\
        Simulation~\footnotemark[1]                               & 28              & 5                                      & 2               & $60 \pm 5$   & $216 \pm 16$    & $0.607 \pm 0.001$                                  \\
                                                                  &                 &                                        & 5               & $50 \pm 5$   & $184 \pm 16$    & $0.604 \pm 0.001$                                  \\
        \cline{3-7}
                                                                  &                 &                                        & 1               & $50 \pm 5$   & $152 \pm 16$    & $0.305 \pm 0.001$                                  \\
                                                                  &                 & 10                                     & 2               & $55 \pm 5$   & $200 \pm 16$    & $0.320 \pm 0.001$                                  \\
                                                                  &                 &                                        & 5               & $60 \pm 5$   & $232 \pm 16$    & $0.321 \pm 0.001$                                  \\
        \hline
        Ref.~\cite{Beare:2020gzr} (S18)~\footnotemark[1]          & 23              & 2                                      & 1               & $85$         & $165$           & $0.547 \pm 0.004$~\footnotemark[2]\footnotemark[3] \\
        \hline
        Ref.~\cite{Beer:2014ooa}~\footnotemark[1]                 & 28              & 5                                      & $4.75 \pm 0.25$ & $30$         & $270$           & $0.265 \pm 0.003$~\footnotemark[2]\footnotemark[3] \\
        \hline
        Ref.~\cite{Antognini:2022ysl} (Aerogel-1)~\footnotemark[1] & 12.5            & 3.4                                    & $4.5 \pm 0.5$   & $45 \pm 5$   & $105 \pm 5$     & $1.22 \pm 0.01$~\footnotemark[2]\footnotemark[4]   \\
        \hline\hline
    \end{tabular}
    \footnotetext[1]{Only statistical errors are shown.}
    \footnotetext[2]{Converted from $R_\text{vac}$ (\cref{tab:maxRvac}) by $Y_\text{vac}=R_\text{vac}f_\text{M}f_{\mu^+}^\text{stop}$, assuming $f_{\mu^+}^\text{stop}=\frac{1}{2}\left(1-\frac{\pi(\phi/2)^2}{\sqrt{3}/2(s+\phi)^2}\right)$ by considering muons are half-stopped in flat target and taking into account the material volume fraction near the surface after ablation, and $f_\text{M}=0.655$~\cite{Beare:2020gzr}.}
    \footnotetext[3]{Includes muonium decays within $10~\text{mm}<z<40~\text{mm}$ only.}
    \footnotetext[4]{Model-dependent assumptions of the temperature at 400~K and the diffusion time at 200~ns.}
\end{table}

With the simulation method validated, we can utilize it to investigate the performance of perforated targets of different ablation configurations to guide the optimization of the mainstream scheme in the related muonium experiment.
In this section, we will study the muonium yield under various target geometries to find the dependencies of muonium yield on key parameters.
The perforation structure is the same as \cref{fig:TargetSchemanticDiagram}, with a target size of $60~\mathrm{mm}\times 60~\mathrm{mm}\times 10~\mathrm{mm}$ and ablation holes with different geometric parameters in the $40~\mathrm{mm}\times 40~\mathrm{mm}$ area at the center of the downstream surface.
These geometric parameters include a hole spacing of 0--100~$\mu$m, a diameter of 40--360~$\mu$m, and depths of 1~mm, 2~mm, and 5~mm.
The temperature is set to 322~K as the same in previous sections.
Three kinds of surface muon beams carrying 28~MeV/$c$ momentum with momentum spread of 2.5\%, 5\%, and 10\% are considered.
In a real experiment, the muon beam usually passes materials such as a beam degrader or a beam monitor~\cite{Xu:2023bml}.
Therefore, we place an aluminum degrader at 5~mm in front of the target where the muon beam goes through and loses energy.
A portion of muons will stop in the target and form muonium atoms.
The degrader with a different thickness will result in various distributions for muon stopping positions. 
Thereby, the degrader can affect the emission of muonium into vacuum and eventually change the vacuum yield.
It means that the degrader thickness should be optimized to achieve the maximum yield, as shown in simulation results \cref{fig:al_degrader_thickness}.
From the results, we infer the optimal degrader thicknesses for beam momentum spread of 2.5\%, 5\%, and 10\% are 430~$\mu$m, 410~$\mu$m, and 370~$\mu$m, respectively.

We use five quantities, $f_{\mu^+}^\text{stop}$, $f_\text{M}$, $Y_\text{tot}$, $R_\text{vac}$, and $Y_\text{vac}$, to characterize the formation and emission of muonium.
They represent the muonium formation fraction, the muon stopping fraction, the muonium total yield, the muonium emission efficiency, and the vacuum muonium yield, respectively.
They are defined as
\begin{equation}
    f_{\mu^+}^\text{stop}=\frac{N_{\mu^+}^\text{stop}}{N_{\mu^+}^\text{in}}~,\quad f_\text{M}=\frac{N_\text{M}^\text{tot}}{N_{\mu^+}^\text{stop}}~,\quad Y_\text{tot}=\frac{N_\text{M}^\text{tot}}{N_{\mu^+}^\text{in}}~,\quad R_\text{vac}=\frac{N_\text{M}^\text{vac}}{N_\text{M}^\text{tot}}~,\quad Y_\text{vac}=\frac{N_\text{M}^\text{vac}}{N_{\mu^+}^\text{in}}~.
\end{equation}
Among them, $N_{\mu^+}^\text{stop}$ is the number of muons stopped in target, $N_{\mu^+}^\text{in}$ is the total number of incident muons, $N_\text{M}^\text{tot}$ is the total number of produced muonium, and $N_\text{M}^\text{vac}$ is the number of vacuum muonium.
There are following notable relationships between them
\begin{equation}
    Y_\text{tot}=f_\text{M}f_{\mu^+}^\text{stop}~,\quad Y_\text{vac}=R_\text{vac}Y_\text{tot}=R_\text{vac}f_\text{M}f_{\mu^+}^\text{stop}~.
\end{equation}
While $f_\text{M}$ is mostly determined by material properties and $f_{\mu^+}^\text{stop}$ is related to both material properties and the beam condition, other three quantities, $Y_\text{tot}$, $R_\text{vac}$, and $Y_\text{vac}$, will be affected by aerogel target geometry.
On one hand, increasing the spacing between holes will increase the total amount of material, thereby increasing the rate of muons stopping in the material and producing muonium, but these muonium atoms are less likely to emit into vacuum due to the material's obstruction.
On the other hand, increasing the diameter of the hole can improve the emission efficiency, but this will reduce the overall muonium yield.
Therefore, it can be expected that there is a geometric parameter combination that leads to the highest vacuum yield when the material and the beam condition are fixed.
The simulation results support this prospect.
As shown in \cref{fig:scan_simulation}, there exists a maximum value for both vacuum yield and emission efficiency.
As the spacing and diameter deviate from the optimal values, the vacuum yield and emission efficiency gradually drop and are significantly suppressed when the spacing or diameter is too tiny.

Based on the notable relation $Y_\text{vac}=R_\text{vac}f_\text{M}f_{\mu^+}^\text{stop}$, the enhancement in muonium vacuum yield $Y_\text{vac}$ can be achieved by three different mechanisms.
\begin{itemize}
    \item Enhancement by sufficient open surface area that enhances muonium diffusion towards vacuum, which contributes positively to $R_\text{vac}$.
    This enhancement is suppressed when the diameter is too small or the spacing is too large, as shown in \cref{fig:scan_simulation}.
    In this case, although muonium atoms are formed near the target surface, there is little opening surface for them to easily diffuse into the vacuum.
    Meanwhile, they have a long distance to travel before leaving the material, during which most of them decay and cannot reach the target surface.
    \item Enhancement by favorable forming position.
    In this case, muoniums originates near the target surface and is favorable for their emission into vacuum, which contributes positively to $R_\text{vac}$.
    This enhancement is suppressed when the spacing is tiny, as shown in \cref{fig:scan_simulation}, or the beam momentum spread is large, as shown in \cref{tab:maxRvac} and \cref{tab:maxYvac}.
    For tiny spacings, muonium tends to form near the bottom of the hole rather than near the target surface, since there is not enough material to support sufficient muonium formation near the surface.
    For large beam momentum spread, the distribution of muonium forming position tends to be uniform, with a large number of muoniums produced deep inside the target.
    Both of these two situations require a long diffusion path for muonium atoms to travel before they emit into vacuum.
    \item Enhancement by appropriate material removal, which contributes positively to $Y_\text{tot}$.
    This mechanism takes effect when materials are not excessively removed (i.e. with large spacing or small diameter, as shown in \cref{fig:scan_simulation}).
    At this point, there are more muons stopped inside the target so the original total yield of muonium is higher, leading to a higher vacuum yield.
\end{itemize}
In conclusion, we prefer to have a high-quality muon beam with minimize momentum spread and an appropriate opening surface area, which can maximize the number of muonium atoms forming near the target surface and maximize the muonium emission into vacuum.
This guides us to optimize the muonium target in the experiment MACE.

\section{\label{sec:summary}Summary and outlook}

We have developed a technique to simulate the muonium production and tracking in perforated silica aerogel, and implemented it in \GeantFour as an independent physical process.
This simulation technique has been validated with the spacetime distribution in muonium decays and the muonium emission efficiency measured in TRIUMF.
The validation has shown a good consistency between simulation and experiment.
In addition, optimization of geometric parameters in the perforated aerogel has been carried out so that we could make good use of the simulation technique to facilitate muonium-related experiments such as MACE.
We have considered a simplified simulation setup based on real experimental circumstances and incident beams under different conditions to investigate the performance of different target geometries.
We have scanned the depth, spacing, and diameter of the holes within a certain range and, for the first time, have determined the explicit dependency of the muonium total yield, the muonium emission efficiency, and the vacuum muonium yield on the geometric parameters.
We have identified the mechanisms for the enhancement of vacuum muonium yield in the experimental design and possible optimization directions are indicated.
Nevertheless, there is still plenty of room for further improvement in the simulation techniques and muonium targets.
On one hand, in the simulation technique, the effect of material removal caused by laser ablation on the muon transportation has not been considered in detail due to the independence of the current process implementation, and the meticulous combination of this muonium tracking process with the muonium conversion process or other muonium-related processes inside the material is still a technical topic needs to be studied.
On the other hand, it is convinced that the combination of a low-momentum-spread muon beam and an optimized target will lead to higher vacuum muonium yields, which deserves further investigation.
We believe that this new simulation method can inspire novel target designs and help achieve better results.
It could be expected that our simulation technique will make an essential contribution to the optimization and guidance of muonium target design and will play an important role in the related science and technology frontiers, which share similar hydrogen-like atom formation and diffusion physics.

\begin{acknowledgments}
We appreciate Dr. Ce Zhang and Prof. Dr. Tsutomu Mibe for fruitful discussions.
We also appreciate Ruixuan Gao for proofreading.
This project was supported in part by National Natural Science Foundation of China under Grant No. 12075326, Natural Science Foundation of Guangzhou under Grant No. 2024A04J6243 and Fundamental Research Funds for the Central Universities (23xkjc017) in Sun Yat-sen University.
The simulation was conducted by a strong support of computing resources from the National Supercomputer Center in Guangzhou.
\end{acknowledgments}

\bibliography{bib}

\end{document}